\documentclass[aps,pre,twocolumn,superscriptaddress]{revtex4}

\usepackage{graphicx,subfigure}
\usepackage{appendix}
\usepackage{amssymb}
\usepackage{amsmath}
\usepackage[british]{babel}
\usepackage{color}

\begin{document}

\title{Non-universal tracer diffusion in crowded media of non-inert obstacles}

\author{Surya K. Ghosh}
\affiliation{Institute for Physics \& Astronomy, University of Potsdam,
14476 Potsdam-Golm, Germany}
\author{Andrey G. Cherstvy}
\affiliation{Institute for Physics \& Astronomy, University of Potsdam,
14476 Potsdam-Golm, Germany}
\author{Ralf Metzler}
\affiliation{Institute for Physics \& Astronomy, University of Potsdam,
14476 Potsdam-Golm, Germany}
\affiliation{Department of Physics, Tampere University of Technology, 33101
Tampere, Finland}

\date{\today}

\begin{abstract}
We study the diffusion of a tracer particle, which moves in continuum space
between a lattice of excluded volume, immobile non-inert obstacles. In particular,
we analyse how the strength of the tracer-obstacle interactions and the volume
occupancy of the crowders alter the diffusive motion of the tracer. From the
details of the partitioning of the tracer diffusion modes between trapping states
when bound to obstacles and bulk diffusion, we examine the degree of localisation
of the tracer in the lattice of crowders. We study the properties of the tracer
diffusion in terms of the ensemble and time averaged mean squared displacements,
the trapping time distributions, the amplitude variation of the time averaged
mean squared displacements,
and the non-Gaussianity parameter of the diffusing tracer. We conclude that
tracer-obstacle adsorption and binding triggers a transient anomalous diffusion.
From a very narrow spread of recorded individual time averaged
trajectories we exclude continuous type random walk processes as the underlying
physical model of the tracer diffusion in our system.
For moderate tracer-crowder attraction the motion is found to be fully ergodic,
while at stronger attraction strength a transient disparity between ensemble and
time averaged mean squared displacements occurs. We also put our results into
perspective with findings from experimental single-particle tracking and
simulations
of the diffusion of tagged tracers in dense crowded suspensions. Our results have
implications for the diffusion, transport, and spreading of chemical components in 
highly crowded environments inside living cells and other structured liquids.
\end{abstract}

\maketitle

\section{Introduction} 

Macromolecular crowding (MMC) abounds in living biological cells, with up to
$\phi\approx30\dots35\%$ of the volume of the cytoplasmic liquid being occupied 
by large biopolymers such as proteins, nucleic acids, ribosomes, as well as
membranous structures, and other complexes \cite{crowd1,crowd2,crowd3,crowd4,
elcock-ecoli-cytoplasm}. These volume-excluding and often non-inert obstacles
alter the diffusion behaviour of cellular components and the rates of biochemical
reactions taking place in this highly complex liquid \cite{ellis01,
crowd-reaction-minton,minton-theory,minton08,kapral12,leduc-nature}. These changes
occur both due to an enhanced solution viscosity \cite{njp14-rings} and the
sheer physical obstruction imposed on particle diffusion due to the presence of
the obstacles. 

The mean squared displacement (MSD) of a tracer particle in such crowded solutions
often becomes anomalous \cite{report,bouchaud,franosch13,pt,pccp,pccp1}
\begin{equation}
\label{msd}
\left<\mathbf{r}^2(t)\right>\sim D_{\beta}t^{\beta},
\end{equation}
where $D_{\beta}$ is the anomalous diffusion coefficient of dimension $\mathrm{
cm}^2/\mathrm{sec}^{\beta}$ and $\beta$ the anomalous diffusion exponent. Its
typical range $0<\beta<1$ indicates slower-than-Brownian, \emph{subdiffusive\/}
motion \cite{report,bouchaud}. Often, the subdiffusion (\ref{msd}) turns out
to be transient and at long times the MSD crosses over to Brownian Motion
\cite{franosch13,sokolov12}. A wide range of scaling exponents $\beta\sim0.4\dots
0.9$ has been reported for the obstructed diffusion of tracers of various sizes and
surface properties in crowded solutions inside cells. Examples include the motion
of small proteins \cite{lang-subdiff-nucleus-fcs}, mRNA molecules \cite{goldingcox},
telomeric chromosomal loci \cite{telosubdiff}, Cajal bodies \cite{cajal-ad}, lipid
and insulin granules \cite{jeon11}, and natural virus particles \cite{brauch01,
brauch02} in the cytoplasm of living cells. Further examples are membrane lipids
and membrane-bound proteins \cite{franosch10,petrov11,rm12lipids}, water molecules
associated to membranes \cite{yamamoto}, hair bundles in ears \cite{hair-subdiff},
as well as several examples for the motion of tracers such as proteins in complex
liquids \cite{fradin05,pan,lene1}. Subdiffusive regimes in crowded systems have
also been observed and modelled for actively driven particles \cite{robert}, and
the dependence of the effective diffusivity $D(\phi)$ reveals a minimum as function
of the MMC volume fraction $\phi$ \cite{ukraine13}.

Anomalous diffusion of the form (\ref{msd}) is modelled in terms of a wide range
of stochastic processes \cite{report,bouchaud,franosch13,sokolov12,pt,pccp,pccp1}.
These include continuous time random walks (CTRW) \cite{montroll,report,bouchaud},
fractional Brownian motion \cite{mandelbrot,deng,pccp,pccp1} and the closely
related fractional Langevin equation motion \cite{lutz,deng,goychuk,pccp,pccp1},
as well as diffusion processes with space \cite{hdp,hdp_pre,massignan,fulinski} and
time \cite{lim,fulinski,thiel} dependent diffusion coefficients. CTRW models are closely related to trap models, in which the tracer is successively
immobilised \cite{ledoussal-diff,kehr-review,bouchaud,eli}. Despite advances in
simulations \cite{poland14-ad,rm12lipids} and theoretical approaches \cite{pccp,
pccp1}, there is no consensus on the physical understanding of the subdiffusion of
passive tracer particles in crowded solutions, that would be directly applicable
to the cytoplasm of living cells \cite{franosch13,pt}. In particular, it is likely
that different
physical origins dominate for different tracer sizes and shapes, as well as length
and time scales of the diffusion. The lack of a consensus picture for diffusion in
MMC environments suggests that the observed anomalous diffusion is not universal
but depends on specific parameters.

To address such specific origins for deviations from Brownian motion, extensive
computer simulations of passive tracer diffusion were performed by several groups.
The recent approaches of Refs.~\cite{berry11,berry13}, for
instance, considering the tracer motion in lattices of immobile, randomly
positioned obstacles or regularly ordered obstacles jiggling in a confining
potential
demonstrated that the anomalous diffusion regime is governed by the obstacle volume
occupancy, with significant subdiffusive motion observed at higher crowder
volume fraction $\phi$. This
subdiffusion is transient and can be quantified by the dependence of the local
anomalous diffusion exponent
\begin{equation} 
\beta(\phi,t)=\frac{d\log\left(\left<\mathrm{\textbf{r}}^2(\phi,t)\right>\right)}{
d\log (t)}
\label{eq-beta}
\end{equation}
and the effective diffusion coefficient $D(\phi,t)$, where we included the explicit
dependencies on the MMC volume fraction $\phi$ and time $t$. In denser obstacle
lattices, diffusion is more localised and the value of $\beta(t)$ smaller
\cite{berry13}. Remarkably, the transient subdiffusion regime was shown to
disappear at higher obstacle mobility \cite{berry11,berry13}. Moreover, the
distribution of particle trapping times in dynamical cages formed between the
crowders observed in simulations for mobile, confined, and static obstacles was
shown to be inconsistent with CTRW models \cite{note-berry,berry11} 
Recent experimental
advances in the field of obstructed obstacle-mediated tracer diffusion, including
the regimes of transient anomalous diffusion, are presented in 
Refs.~\cite{polska14,weiss09-fbm,weiss14-pccp}.

The study of obstructed diffusion by means of simulations was pioneered by Saxton
in a series of papers \cite{saxton-1,saxton-binding-ad,saxton-3}. In particular,
the effects of the fraction of lattice sites occupied by crowders and of their
diffusivity were examined. Specifically,
effects of
tracer-obstacle binding on the anomalous diffusion properties were studied and
connected to a binding energy landscape for immobile point-like obstacles
positioned at a fixed concentration on the lattice \cite{saxton-binding-ad}.
Obstructed diffusion of point-like tracers in a lattice of randomly
positioned, static obstacles was investigated in Ref.~\cite{netz-dorf} and
shown to give rise to a reduction of the tracer diffusivity $D$ with the obstacle
concentration. Nonzero values of the diffusivity $D$ even for very densely packed
obstacles appear in such a model due to the existence of a percolation structure. 
True long-time subdiffusion can only be realised at the percolation threshold in
such lattices of randomly distributed obstacles \cite{franosch13}. On a cubic
lattice, the critical percolation thresholds corresponds to 31\% \cite{franosch13}.
For a random walker on the infinite incipient cluster, the scaling exponent of the
MSD is $\beta\approx0.697$ \cite{yasmine}. The physical reason is the formation of
a labyrinthine-like environment \cite{sokolov12}, in which the tracer needs to
escape dead ends and cross narrow causeways present on all scales. For lattices
below the percolation transition as well as for regularly positioned obstacles,  
as in the current study, the anomalous diffusion regime is transient.

An interesting alternative to the modelling of transient anomalous diffusion are
Lorentz gas-based models which were developed to exploit the localisation
transitions on a percolation network of overlapping spherical obstacles
\cite{fran06,fran11}. The scaling relations for the suppression of the
tracer diffusivity as the system approaches the critical percolation density $\bar
\phi$ was determined, namely $D(\phi)\sim[(\phi-\bar\phi)/\bar\phi]^\mu$, 
where the percolation exponent is $\mu\approx 2.88$ \cite{fran06}. At the
percolation threshold, persistent anomalous diffusion with exponent
$\beta=2/6.25\approx0.32$
was found, for even denser systems the particles are eventually localised
\cite{fran06}.

\begin{figure*}
\includegraphics[height=18cm,angle=270]{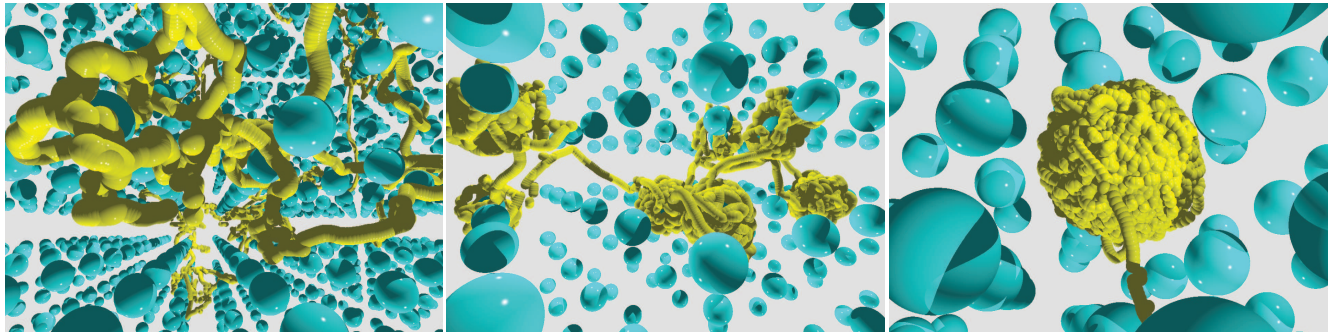}
\caption{Schematic of the three-dimensional tracer-obstacle system used in
our simulations, for the obstacle radius $R=0.6$ and tracer-obstacle binding
strength $\epsilon_A=2$, $6$, and $10$ (from left to right). The
tracer trajectories, as obtained directly from the simulations, are of the same
length in all the three panels. Note that the particle traces are rendered with
a finite thickness although the tracer particle is point-like. The fraction of
time that the particle spends in the surface-bound diffusion mode grows with
$\epsilon_A$.} 
\label{fig-scheme}
\end{figure*}

Here, we extend the class of systems considered by Saxton \cite{saxton-binding-ad} 
based on transient binding of tracer particles to physical obstacles. We perform
extensive Monte-Carlo simulations of tracer diffusion on 3D lattices of sticky
spherical obstacles of varying radius $R$.
In this obstruction-binding diffusion model we examine the trapping time 
distributions of the tracers, the time averaged MSD --- which is a more relevant
observable when compared to experimental situations than the ensemble averaged
MSD (\ref{msd}) --- and the effective tracer diffusivity $D$. The model parameters
are systematically varied, including the crowder radius $R$ and thus the volume
fraction of crowders $\phi$ and the tracer-crowder binding energy $\epsilon_A$.
A schematic of the system is shown in Fig.~\ref{fig-scheme} along with sample
trajectories of a tracer particle. These novel features substantially extend the 
known simulations results for obstructed tracer diffusion on 2D lattices of
reflecting spherical \cite{berry11,berry13} and cylindrical \cite{bezrukov12} obstacles.
Despite the difference of mobile polymer obstacles to our scenario of ordered
reactive crowders, our results show interesting similarities with the diffusion
of tracer particles in dense solutions of non-inert polymer chains recently
reported by the Holm group \cite{holm13-cylinder}. We will discuss the
consequences of this similarity below.

The paper is organised as follows. In Section \ref{sec-model} we introduce the
basic notations and the quantities to be analysed. We outline the computational
scheme and theoretical concepts. In Section \ref{sec-results} we report the main
simulations results and support them by theoretical scaling arguments. We analyse
the effects of the MMC volume fraction and the strength of the obstacle-tracer
binding. Moreover, we compute the ensemble and time averaged particle
displacements as well as the distributions of particle trapping times to the
sticky obstacles. To rationalise the stochastic behaviour and determine the
concrete underlying effective diffusion model, we also systematically compute the 
non-Gaussianity parameter $G$. In Section \ref{sec-discussion} the conclusions
are drawn and possible applications of our results to some experimental systems
discussed.

\section{Simulation model and approximations}
\label{sec-model}

To mimic the conditions of a crowded environment, we consider a primitive cubic
lattice every site of which is occupied by a spherical obstacle, as shown in
Fig.~\ref{fig-scheme}. The maximal size of the obstacle $R_{\mathrm{max}}$ for the
conditions of  close packing is $R_{\mathrm{max}}=a/2$ where $a$ is the lattice
constant. In the following we will use $a=2$ in dimensionless units. The maximal
volume occupancy by obstacles on such a static cubic lattice is $\phi_{\mathrm{
max}}=\pi/6\approx0.524$, compared to $\overline{\phi}_{\mathrm{max}}=\pi/\sqrt{18}
\approx0.740$ for the densest packing of spheres in 3D
\cite{greek-dense}. The obstacles are
considered immobile in our simulations.

In the simulations presented below, the point-like tracer starts
in the centre of a cage, at the maximal
distance from the eight surrounding obstacles. At the very start the tracer
particle thus performs free motion, until it encounters the surface of a
crowding particle to which it can subsequently bind. The length scale of the
spatial heterogeneity $l^\star$ in the system is of the order of the free path
of the tracer between neighbouring obstacles, $l^{\star}(\phi)\sim a(1-\phi^{1/3})
\sim\sqrt{D_0t^\star}$. As we will show such heterogeneities effect subdiffusion
at intermediate time scales $t^\star$, while at much longer time-scales the
diffusion becomes Brownian, as demonstrated in Fig.~\ref{fig-msd}. As many
binding-unbinding events take place during the length of the recorded traces
in our simulations, the initial particle position does not affect the long time
dynamics. The tracer particle becomes adsorbed onto the obstacle surface with the
binding energy $\epsilon_A$ and stays in the bound state for the
average adsorption time $t_{\mathrm{ads},i}$. While bound, the tracer diffuses
along the spherical obstacle surface with the same diffusion coefficient as in
the free unbound state, that is, it moves along the surface of a crowder sphere
with $D_{\mathrm{ads}}=D_0$. The tracer is considered unbound once it separates
from the obstacle for more than the distance $0.1R_{\mathrm{max}}$, see also the
definition of the interaction potential
below.\footnote{To map the exchange between the bulk and a
reactive surface (here defined via the attractive part of the potential
(\ref{lj_pot}) onto a random walk picture, compare Refs.~\cite{irwin}.} In our
simulations we place a single tracer on the crowder lattice and then average over
many individual traces. The repeated binding and unbinding events separating the
particle motion between surface and bulk diffusion lead to an effective
distribution between the modes of tracer motion, which is reflected in the tracer
particle MSD and other diffusive characteristics such as average trapping times,
see below.

The attractive interactions between the mobile tracer particle and immobile crowder
spheres are modelled in terms of the Lennard-Jones 6-12 potential,
whose attractive branch is cut off at the distance $r_\text{cutoff}$,
namely
\begin{equation}
\label{lj_pot}
U_{\mathrm{LJ}}(r)=\left\{\begin{array}{ll}
4\epsilon_A\left[\left(\frac{\sigma}{r}\right)^{12}-\left(\frac{\sigma}{
r}\right)^6\right]+\epsilon_{\mathrm{cutoff}}, & r\le r_\text{cutoff}\\[0.32cm]
0, & r>r_\text{cutoff}
\end{array}\right..
\end{equation}
The parameter $\sigma$ is connected to the obstacle radius by $R=2^{1/6}\sigma$
corresponding to the minimum of $U_{\mathrm{LJ}}(r)$. The attractive potential
$U_{\mathrm{LJ}}(r)$ is truncated at the critical distance $r_
{\text{cutoff}}=R+
0.1R_{\mathrm{max}}$ thus mimicking an attractive shell of a fixed width $0.1R_{\mathrm{max}}$
around the obstacles. Thus, the thickness of the attractive layer
around obstacles of different sizes is the same. The vertical
energy shift $\epsilon_{\mathrm{
cutoff}}$ is a constant  which sets  $U_{\mathrm{LJ}}(r_\text{cutoff})=0$.

We simulate the motion of the point-like tracer of mass $m$  with coordinate $\mathbf{r}(t)$
in the presence of friction based on the Langevin equation
\begin{equation}
m\frac{d^2\mathbf{r}(t)}{dt^2}=-\gamma\frac{d\mathbf{r}(t)}{dt}+\boldsymbol{\xi}
(t)-\boldsymbol\nabla\sum_jU_{\mathrm{LJ}}(|\mathbf{r}-\mathbf{R}_j|).
\label{eq-langevin}
\end{equation}
Here $\gamma$ is the friction coefficient coupled to the strength of the Gaussian
noise through
\begin{equation}
\left<\boldsymbol{\xi}(t_1)\cdot\boldsymbol{\xi}(t_2)\right>=6k_B\mathcal{T}\gamma m
\delta(t_1-t_2),
\end{equation}
where $\delta(\cdot)$ denotes the Dirac delta-function and
$k_B\mathcal{T}$ represents the thermal energy. The noise has zero mean, and
has vanishing correlations in the different Cartesian directions. The sum in
Eq.~(\ref{eq-langevin}) runs over the positions of all crowding particles
$\textrm{\textbf{R}}_j$. The fact that we consider a point-like particle is not
a severe restriction, as a finite size of the tracer particle would correspond
to a re-normalisation of the crowder radius, compare also Ref.~\cite{netz-dorf}.
Concurrently
the surface diffusivity of the tracer along the crowders would need to be adjusted.

In our simulations we neglect tracer-obstacle hydrodynamic interactions, which can,
in principle, affect the long-time behaviour of the system. In particular, because 
of their long-range $1/r$-nature \cite{kapral12}, the diffusing particles can feel
the obstacles at a finite distance without direct collisions, see e.g. Ref.~\cite{ukraine13}. Note however that hydrodynamic interactions have
recently also been demonstrated to affect the short-time tracer diffusion
dynamics in fluids \cite{experHDI,greb13HDI}.

\begin{figure}
\includegraphics[width=8cm]{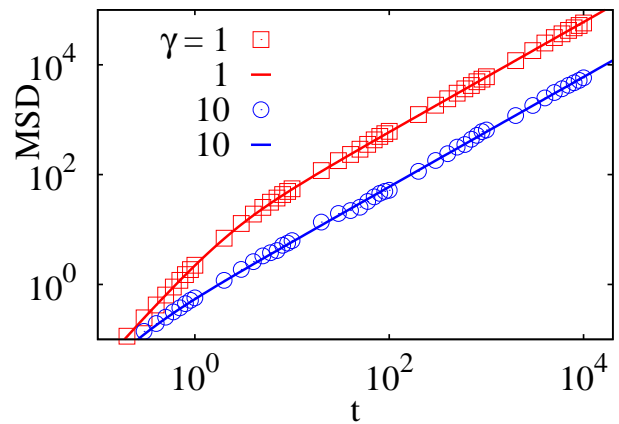}
\caption{MSD from simulations (dots) of the free underdamped Langevin equation
versus the analytical Ornstein-Uhlenbeck solution (\ref{eq-ou}) (solid curves),
plotted for different values of the friction coefficient $\gamma$, as indicated
in the plot. The ballistic regime can be clearly distinguished in the case of lower friction.} 
\label{fig-free-diff}
\end{figure}

\begin{figure*}
\includegraphics[width=14cm]{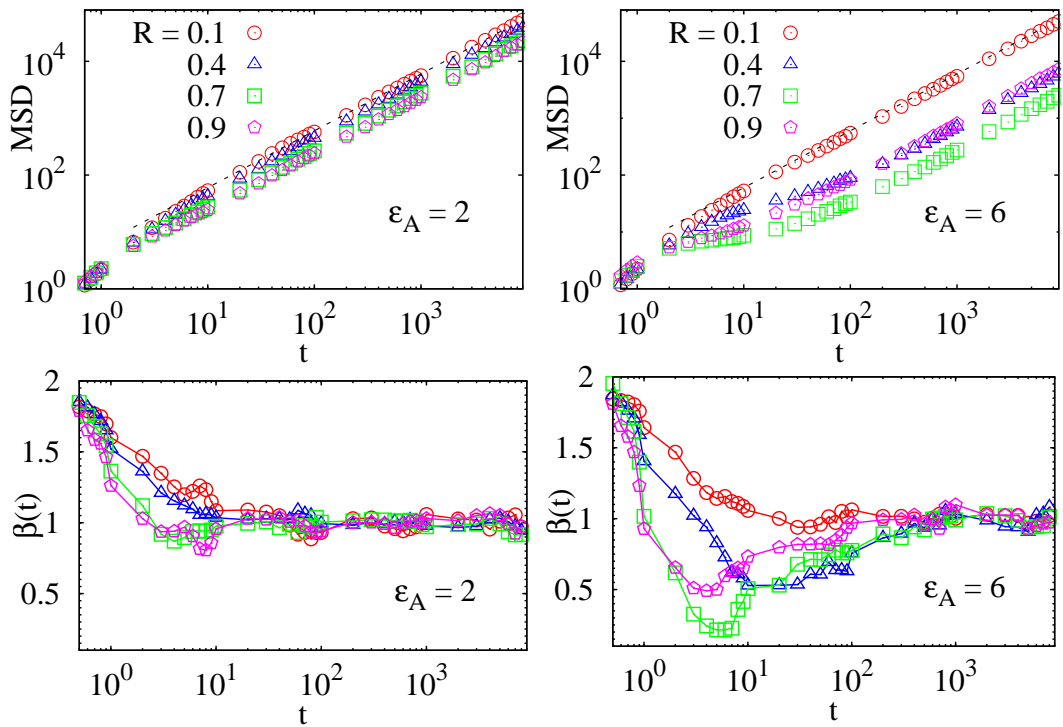}
\caption{MSD $\langle\mathbf{r}^2(t)\rangle$ of the tracer particles and the 
corresponding scaling exponent $\beta(t)$, plotted for two trace-crowder
attraction strengths, $\epsilon_A=2$ and $6$, in units of $k_B\mathcal{T}$.
The result (\ref{eq-msd-bm}) for the MSD of three dimensional Brownian motion
is represented by the dashed line in the MSD plots. The number of trajectories
used for the averaging is $N=10^3$, and each trajectory consists of $10^7$
steps corresponding to the length $T=10^4$ of the simulated time series $\mathbf{
r}(t)$ in terms of the simulation step $\delta t=10^{-3}$. Other parameters are as indicated in the plots. The obstacle radius is given in terms of $R_{\text{max}}=a/2$. The data sets for varying crowder radius are shown by open symbols.}
\label{fig-msd}
\end{figure*}

In free space the solution of the Ornstein-Uhlenbeck process (\ref{eq-langevin})
without the last term is well known \cite{coffey},
\begin{equation}
\left<\textrm{\textbf{r}}^2(t)\right>=\frac{6mk_B\mathcal{T}}{\gamma^2}\left[
\frac{\gamma t}{m}-1+\exp\left(-\frac{\gamma t}{m}\right)\right]
\label{eq-ou}
\end{equation}
describing the crossover from initial ballistic motion
\begin{equation}
\label{msd_ball}
\left<\textrm{\textbf{r}}^2(t)\right>\sim(3k_B\mathcal{T}/m)t^2
\end{equation}
before the characteristic time $m/\gamma$ to overdamped, Brownian motion
\begin{equation}
\label{eq-msd-bm}
\left<\textrm{\textbf{r}}^2(t)\right>\sim(6k_B\mathcal{T}/{\gamma})t.
\end{equation}
This solution is reproduced by our simulations, see Fig.~\ref{fig-free-diff}.
The local scaling exponent $\beta(t)$ is computed as
discretised logarithmic derivative from the MSD traces obtained from the
simulations (i.e., we compute the local derivative of $\log(\mathrm{MSD})$ with
respect to the logarithmically sampled time, see also Eq. (3) of
Ref.~\cite{weiss14-pccp}).

The particle mass is  $m=1$ throughout the paper (we made sure
that the code works fine for varying particle mass and solution friction, as shown
in Fig.~\ref{fig-free-diff}). In
all figures below, time $t$ is shown in units of the simulation step $\delta
t=0.001$ of the Verlet velocity integration scheme, the displacements appear in
units of the lattice constant $a$.
The obstacle size below is given in terms of the maximal
geometrically allowed radius
\begin{equation}
R_\text{max}=a/2
\end{equation}
on the square lattice. Note that even zero-sized crowders $R=0$ have an
attractive shell of finite width around them, as determined by the specific nature
of the attractive potential (\ref{lj_pot}). In the text, however, when we talk
about "free diffusion", no crowders are included in the simulations at all.
In the presence of the sticky, excluded volume obstacles
an exact solution is not known, and we thus analyse this case by simulations. We
find that the fraction of time that the tracer particle spends in the surface-bound
mode increases with the volume occupancy by obstacles and with the tracer-obstacle
affinity $\epsilon_A$. The effect of MMC on the long-time particle diffusivity
$D(\phi)$ reveals a non-trivial dependence at larger $\epsilon_A$, as shown below.

\section{Results}
\label{sec-results}

We discuss the simulations results with respect to three main quantities. In
section \ref{sec_msd} we study the ensemble averaged MSD and the associated
effective diffusivity. The trapping times spent by the tracer on the obstacle
surface are analysed in section \ref{sec_trap}, and in section \ref{sec_tamsd}
we investigate the time averaged MSD based on single trajectory position time
series.

\subsection{MSD, scaling exponent, and effective diffusivity}
\label{sec_msd}

\begin{figure}
\includegraphics[width=8cm]{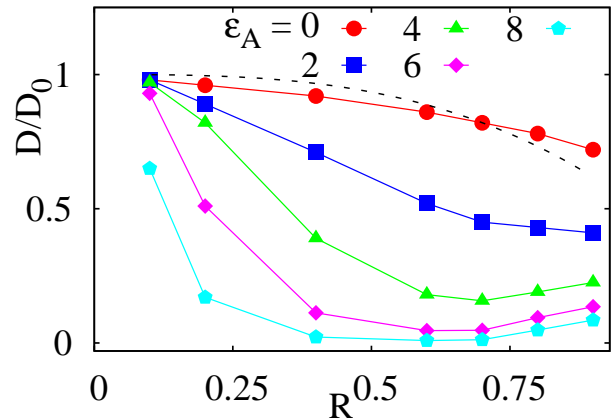}
\caption{Long-time diffusivity $D(\phi)/D_0$ in the terminal Brownian regime as
function of the obstacle size, plotted for varying tracer-obstacle attraction
strengths. Other parameters are the same as in Fig.~\ref{fig-msd}. The dotted
line represents the prediction by  Eq.~(\ref{D-phi-kehr}) for inert obstacles. 
The data sets for varying adsorption strength are shown as filled symbols.}
\label{fig-dc-ratio}
\end{figure}

We study the MSD $\langle\mathbf{r}^2(t)\rangle$ of the tracer particles at
varying volume occupancy $\phi$ of obstacles and the tracer-obstacle affinity
$\epsilon_A$. The main results are shown in Figs.~\ref{fig-msd} and
\ref{fig-dc-ratio}. Generally, we observe that for small interaction strengths
$\epsilon_A$ the anomalous diffusion exponent $\beta(t)$ varies along the time
evolution of the MSD $\langle\mathbf{r}^2(t)\rangle$. At short times it starts
out with the underdamped ballistic motion (\ref{msd_ball}) of the above
Ornstein-Uhlenbeck process, as demonstrated in Fig.~\ref{fig-free-diff}. Such a
short time superdiffusive behaviour is due to inertial effects and was indeed
observed experimentally, for instance, in single particle tracking studies of 
fluorescent beads in sucrose solutions \cite{weiss14-pccp}; see also the detailed studies of Refs. \cite{experHDI,greb13HDI} of inertial effects for the particle  diffusion in a fluid. Subsequently,
Fig.~\ref{fig-msd} demonstrates that the
MSD crosses over to a transient subdiffusive regime with $0<\beta(t)<1$. In the
long time limit the tracer particle performs Brownian motion with a linear scaling
of the MSD and $\beta(t)=1$. Concurrently this Brownian motion regime is
characterised by a reduced diffusivity $D(\phi)$ as compared to unrestricted
Brownian motion of the tracer. The dependence of the effective diffusivity
$D(\phi)$ on the MMC volume fraction $\phi$ is shown for different interaction
strengths in Fig.~\ref{fig-dc-ratio}. For weakly adhesive obstacles, the diffusion
becomes monotonically slower for larger crowders positioned on the lattice, see
Fig.~\ref{fig-dc-ratio}.

Let us look at these behaviours in some more detail. The variation of the scaling
exponent $\beta(t)$ with the crowding fraction and tracer-crowder attraction
strength is shown in Fig.~\ref{fig-msd}. We observe that as both $\epsilon_A$ and
$\phi$ increase, the transient subdiffusion regime progressively extends over a
larger time window. Remarkably, this anomalous diffusion spans up to two decades
of time for substantial tracer-obstacle attraction strengths, see the plot of
$\beta(t)$ for $\epsilon_A=6k_B\mathcal{T}$ in Fig.~\ref{fig-msd}. For relatively weak tracer-obstacle attraction, at $\epsilon_A=2 k_B\mathcal{T}$, only marginally anomalous tracer diffusion was detected, with $\beta\sim 0.85 \dots 1$. Overall, the
subdiffusion regime extends over one to two decades of time, similar to the results
for inert static, randomly-positioned obstacles \cite{netz-dorf} or for the motion
of tracers in a random channel with sticky surfaces \cite{max}.

A physically similar renormalisation of the particle
diffusivity was discovered in Ref.~\cite{poon09pnas} for glassy states in
sticky-particle systems at relatively large volume fractions $\phi$. The
phase diagram of the hard sphere mixture with a short range inter-particle
attraction as well as a self-diffusive MSD dependence were examined, for instance,
by simulations in Ref.~\cite{poon09pnas}. The implications of the square-well
sphere-sphere interaction potential $\epsilon_A$ and volume fraction $\phi$
of crowders were rationalised in detail. A progressive slowing down of the
particle self-diffusion in the attractive hard-sphere mixtures as functions of
$\phi$ and $\epsilon_A$ observed in Figs.~4 and 5 of Ref.~\cite{poon09pnas}
is similar to the properties of the tracer diffusion on the lattice of
moderately-sticky crowders examined here.

For non-attracting obstacles corresponding to $\epsilon_A=0$ the ratio $D(\phi)/
D_0$ of the diffusing coefficient of the tracer particle versus its value in an
un-obstructed environment as function of the obstacle volume occupancy $\phi$ in
a simple mean field approach is predicted to scale as \cite{kehr-review} 
\begin{equation}
\frac{D(\phi)}{D_0}\sim1-\phi=1-\pi\frac{x^3}{6},
\label{D-phi-kehr}
\end{equation}
based on a volume exclusion argument in section 8.3 of Ref.~\cite{kehr-review}. 
Here $x=2R/a$ is the relative obstacle size with respect to the lattice spacing
as used in simulations. The reader is also referred to Ref. \cite{wade-crowder-shape-BJ} for the next-order corrections in the dependence of $D(\phi)$ on the volume fraction of crowders. Indeed, in absence of an attraction between the tracer
and the obstacle surfaces, the behaviour of $D(\phi)/D_0$ as function of $R$
shown in Fig.~\ref{fig-dc-ratio} is in quite good agreement with the prediction
(\ref{D-phi-kehr}). 

For moderate attraction, $\epsilon_A=2k_B\mathcal{T}$, the
decrease of $D(\phi)/D_0$ with $R$ becomes less pronounced at $R$ values larger than
$R\approx0.6$. For even stronger tracer-obstacle interaction, $\epsilon_A\ge4k_B
\mathcal{T}$, remarkably the long-time diffusivity $D(\phi)$ becomes
\emph{non-monotonic\/} with the crowder size
$R$, as shown in Fig.~\ref{fig-dc-ratio}. Physically, at higher obstacle volume
fraction $\phi$ the available space for tracer diffusion becomes effectively
reduced from the three dimensional volume to a lower dimensional space. This
creates pathways or channels between the ``cages'' created by the obstacle and
can effectively speed up the exploration of space by the tracer particle at
higher $\phi$ fractions. Note that this effect would be modified when the surface
diffusivity were considerably smaller than the volume diffusivity. However, as
shown in the discussion of the trapping times below, another contribution to
this speed up-effect could be that for larger crowder radius $R$ the tracer is
in a limbo between vicinal attractive surfaces and thus manoeuvres between
obstacles without binding to them.

We note that in Ref.~\cite{bezrukov12} the diffusion coefficient of a tracer on
an array of cylindrical obstacles on a static, two dimensional square lattice was 
analysed in terms of the generalised Fick-Jacobs equation and by Brownian dynamics
simulations. For the relative diffusivity as function of the crowding fraction
$\phi$ the analogous behaviour $D(\phi)/D_0=1-\phi=1-\pi(R/R_{\mathrm{max}})^2/4$
was found, where $\pi/4$ is the maximal surface coverage in this situation
\cite{bezrukov12}.

\subsection{Statistics of tracer trapping times}
\label{sec_trap}

\begin{figure}
\includegraphics[width=8.5cm]{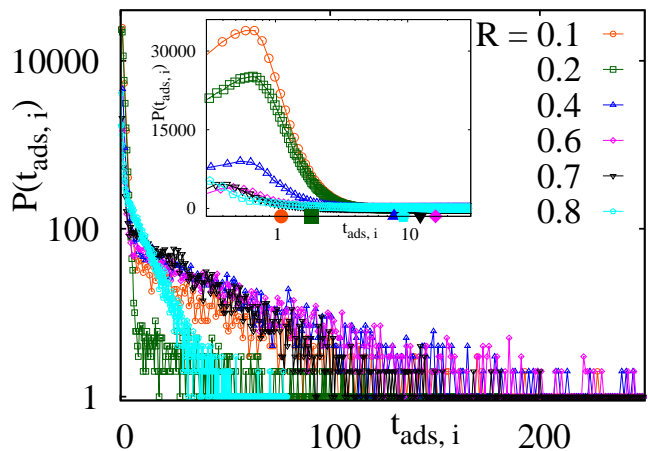}
\caption{Histograms $P(t_{\mathrm{ads},i})$ of individual binding times $t_{
\mathrm{ads},i}$ spent in the obstacle-adsorbed state by the tracer particles,
plotted  for the parameters of Fig.~\ref{fig-msd} with
attraction strength $\epsilon_A=6k_B\mathcal{T}$. The mean values $\left<t_{
\mathrm{ads}}\right>$ for each crowder radius are indicated as the dots on the
horizontal axis of the inset with the corresponding colour, indicating a
non-monotonic behaviour with $R$. The main graph is shown in log-linear scale
to show the exponential tails of the distributions, whereas
the inset features a linear-log scale. The total length of the simulated
trajectories is $10^7$ steps. The obstacle radius $R$ is given in the legend
in terms of $R_{\text{max}}=a/2$.}
\label{fig-hist}
\end{figure}

The non-monotonic dependence of the long-time tracer particle diffusivity on the
crowding fraction is also manifested in the non-monotonic variation of the times
that the tracer spends in the obstacle-adsorbed state. In Fig.~\ref{fig-hist} we
present the statistics of individual adsorption times to the crowding particles
along a very long trajectory containing many binding-unbinding events. In the main
graph of Fig.~\ref{fig-hist} we observe the distribution of tracer-obstacle adsorption times features a peak at short $t_{ads}$, while the
tails of the histograms indicates the expected exponential decay. The corresponding mean adsorption time $\left<t_{\mathrm{ads}}\right>$ is the average over all the binding events encountered in our simulations for a particular set of the model parameters. As we can see,
for larger crowders these exponential tails become progressively longer with
increasing obstacle size up to some $R<0.6\times a/2$, that is, the duration of
binding events can become
significantly longer. In the inset of Fig.~\ref{fig-hist} we use a logarithmic
abscissa to pronounce the initial peak of the histograms. From this plot it
becomes clear that several extremely long binding events can shift the mean
$\left<t_{\mathrm{ads}}\right>$ of the binding time significantly with respect to
the most likely value, compare the relative positioning of the maximum of the
histograms and the mean values indicated by the coloured dots. 

Thus, a small number of extremely long binding events govern the corresponding mean adsorption time $\left<t_{\mathrm{ads}}\right>$. Here, we mention the related study of Ref.~\cite{greb12forcedtocite} in which the modes of surface versus bulk diffusion of a tracer in spherical domains were investigated. Also note that a \textit{different} tracer diffusivity in the obstacle-bound mode as compared to the bulk diffusion can give rise to new interesting effects. In particular, an \textit{optimisation }of the overall passage  times of a  tracer in the target-search problems on a lattice of trapping sites (obstacles) with a likely slower diffusivity should be analysed in the future.

\begin{figure*}
\includegraphics[width=8cm]{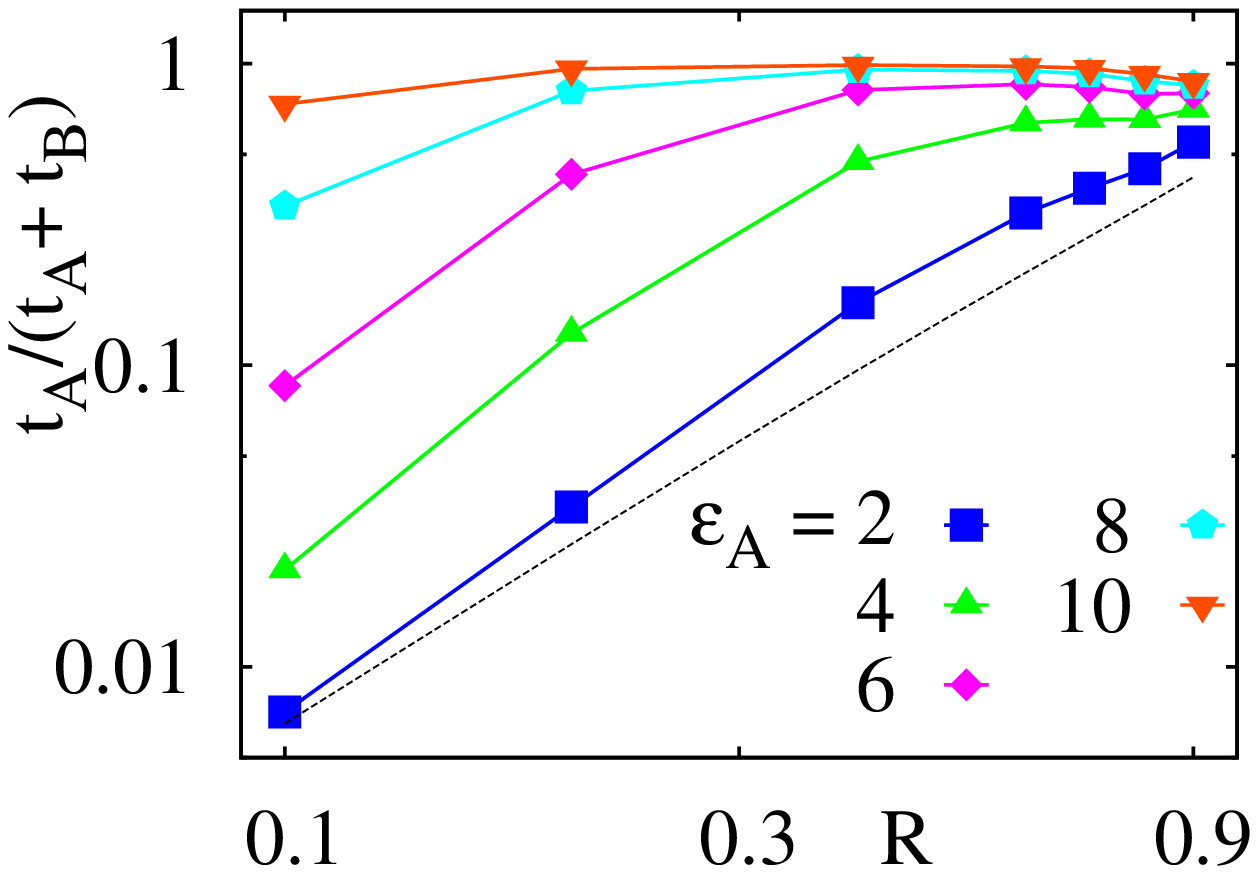}
\hspace*{0.4cm}
\includegraphics[width=7.8cm]{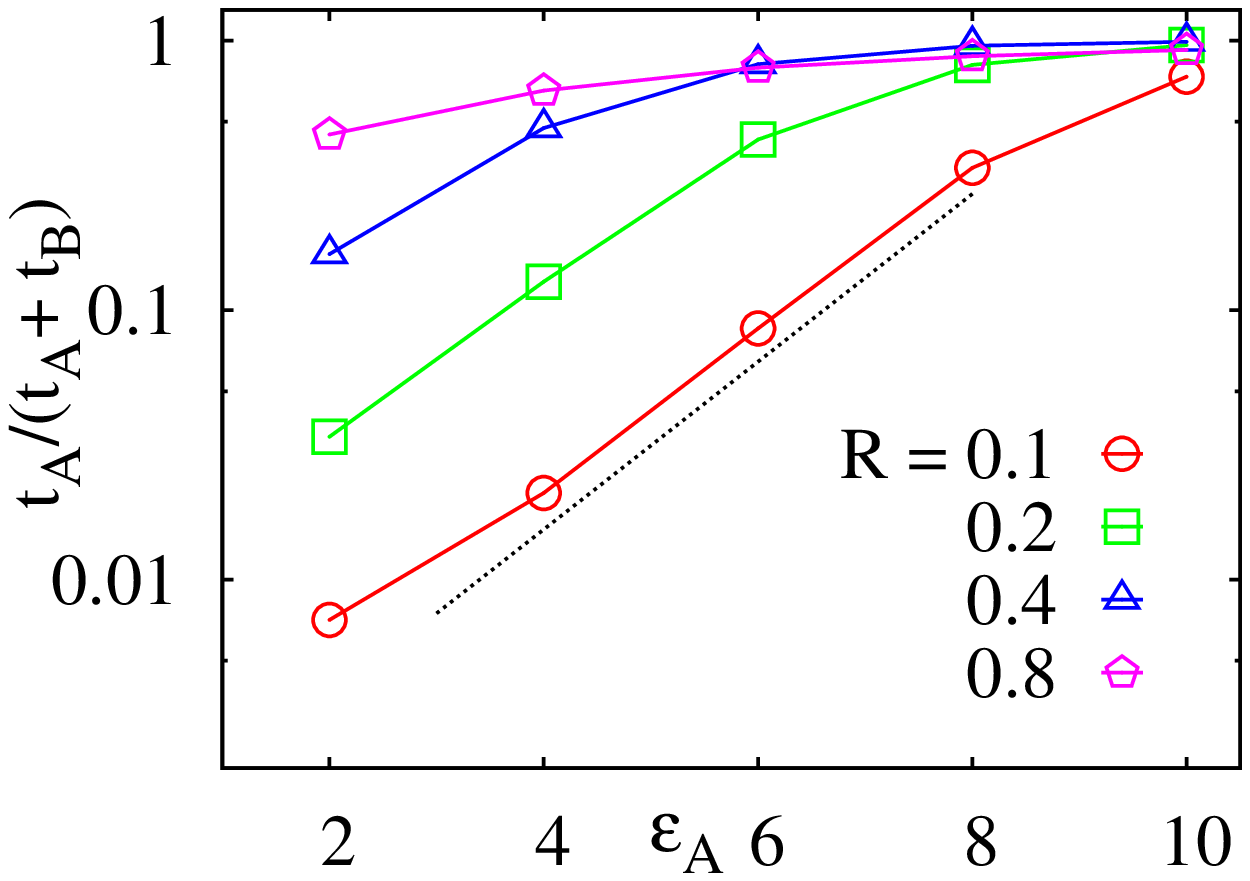}
\caption{Left: total diffusion time $t_A$ in the adsorbed state of the tracer to the
obstacle surface versus the obstacle size $R$, plotted in the log-log scale for varying binding
strengths. Data obtained from averages of the histograms
as those presented in Fig.~\ref{fig-hist}. The dotted asymptote represents
Eq.~(\ref{eq-fraction}). Right: plot of $t_A$ versus the tracer-binding
attraction strength $\epsilon_A$ for varying obstacle radius. The dotted line indicates the exponential
activation of $t_A$ mentioned in the text. At higher $\epsilon_A$ a saturation of $t_A$ is
observed. This saturation is reached for smaller $\epsilon_A$ values when the
obstacle size $R$ is increased.}
\label{fig-ads-time-ave}
\end{figure*}

Let us now turn to the \textit{total} time of adsorption $t_A$ experience by the tracer
particle during a trajectory of duration $T$ generated. We thus sum up all the
adsorption times experienced by the tracer,
\begin{equation}
t_A=\sum_it_{\mathrm{ads},i}.
\end{equation}
Adding to this quantity the excursion times in the bulk between the obstacles, 
\begin{equation}
t_B=\sum_it_{\mathrm{bulk},i},
\end{equation}
we have $T=t_A+t_B$ (the subscripts $A$ and $B$ denote the quantities in the adsorbed and bulk state, respectively). As expected, the total adsorption time $t_A$ obtained from
the simulations is an increasing function of the obstacle size $R$, compare
Fig.~\ref{fig-ads-time-ave}. The initial increase of $t_A$ with the crowder size
$R$ can be understood in terms of the concept of a ``phantom sphere'. Namely, the surface
of a crowding sphere available for surface diffusion by the crowding particles is
$S(R)=4\pi R^2$. If the remaining volume $a^3-{4}\pi R^3/3$ in a cubic unit
cell framed by eight crowding spheres were converted into a (``phantom'') sphere,
the surface of that sphere would amount to
\begin{equation}
S_{\mathrm{ph}}(R)=4\pi\left[\frac{6}{\pi}\left(\frac{a}{2}\right)^3-R^3\right]^{
2/3}. 
\end{equation}
Following this crude argument to simply divide up the time spent on the surface
of the crowders and on the phantom sphere surface corresponding to the free
volume of a unit cell in the obstacle lattice, the ratio of the adsorption time
to the total time $T$ of a sufficiently long trajectory can then be written as 
\begin{equation}
\frac{t_A}{t_A+t_B}\sim\frac{S}{S+S_{\mathrm{ph}}}=\frac{x^2}{x^2+[6/\pi-x^3]^{
2/3}},
\label{eq-fraction} 
\end{equation}
where we again used the dimensionless variable $x=R/R_{\text{max}}=2R/a$. The maximal volume of the phantom sphere is given by the volume $a^3$
of the unit cell, corresponding to an effective radius of the phantom sphere of
$R_{\mathrm{max,ph}}=(a/2)(6/\pi)^{1/3}$. The minimal effective radius of the phantom sphere corresponds to obstacles, which just touch each other on
the square lattice, namely $R_{\mathrm{min,ph}}=(a/2)(6/\pi-1)^{1/3}$.

The ratio $t_A/T$ given by
Eq.~(\ref{eq-fraction}) is quadratic in the obstacle size $R$ for small volume
occupancy by obstacles. This result is in agreement with our simulations results
for moderate tracer-obstacle binding, as shown in Fig.~\ref{fig-ads-time-ave}. In
fact, for the attraction strength $\epsilon_A=2k_B\mathcal{T}$ the agreement with
result (\ref{eq-fraction}) is quite good given the simplicity of our model. In the shown double-logarithmic scale
of Fig.~\ref{fig-ads-time-ave} over the range $R/R_{\text{max}}=0.1\ldots0.9$ the law
(\ref{eq-fraction}) in fact only weakly deviates from the quadratic scaling $t_A/(t_A+
t_B)\sim(\pi/6)^{2/3}x^2$. Naturally, the data for increasing binding affinity
progressively deviate from the formula (\ref{eq-fraction}), for the strongest 
binding strength shown the particles move almost exclusively on the obstacle
surface, independently of the obstacle size.

We observe that for small obstacle sizes and weak tracer-obstacle attraction
strengths the total adsorption time $t_A$ at fixed $R$ grows exponentially
with $\epsilon_A$, $t_A\simeq\exp\left({|\epsilon_A|}/[{k_B\mathcal{T}]}
\right)$, corresponding to the Boltzmann activation in an equilibrium system.
As demonstrated in Fig.~\ref{fig-ads-time-ave} on the right,
for strong tracer-crowder attraction a saturation in $t_A(\epsilon_A)$ is reached, 
such that the activation curve for the ratio $t_A/(t_A+t_B)$ is analogous to the
expression for a simple two level system, $t_A/(t_A+t_B)\sim\exp(|\epsilon_A|/[k_B
\mathcal{T}])/\{1+\exp(|\epsilon_A|/[k_B\mathcal{T}])\}$. 
For larger crowders, when the tracers are confined predominantly to the obstacle
surface, the saturation effect at larger attraction strengths is more pronounced,
see Fig.~\ref{fig-ads-time-ave}. 

In agreement with the results of Fig.~\ref{fig-msd} for the MSD, at moderate
tracer-obstacle binding the adsorption time progressively increases for more 
voluminous obstacles on the lattice. In contrast, at strong tracer-crowder
adsorption the adsorption time initially increases with the obstacle size
\begin{equation}
R=a[{3\phi}/({4\pi)}]^{1/3}
\end{equation}
but above a critical volume fraction $\phi$ the value of $t_A$ \textit{decreases},
as seen in Fig.~\ref{fig-ads-time-ave} for the values $\epsilon_A=8k_B\mathcal{T}$
and $10k_B\mathcal{T}$, as well as for the mean values shown in Fig.~\ref{fig-hist}.
This decrease of the mean adsorption time and thus a stronger contribution of 
bulk excursions is consistent with the non-monotonic dependence of $D(\phi)/D_0$
at large tracer-obstacle binding strengths $\epsilon_A$ and with the enhanced
diffusivity for larger strongly adhesive obstacles, see the curve for $\epsilon_A
=8k_B\mathcal{T}$ in Fig.~\ref{fig-dc-ratio}. We ascribe this small yet somewhat
counterintuitive effect to a competition of binding to neighbouring surfaces
due to which the tracer particle is in a limbo in the bulk, possibly in
conjunction with the reduced effective dimensionality of the environment mentioned
above.

\subsection{Time averaged MSD and non-Gaussianity parameter}  
\label{sec_tamsd}

\begin{figure*}
\includegraphics[width=16cm]{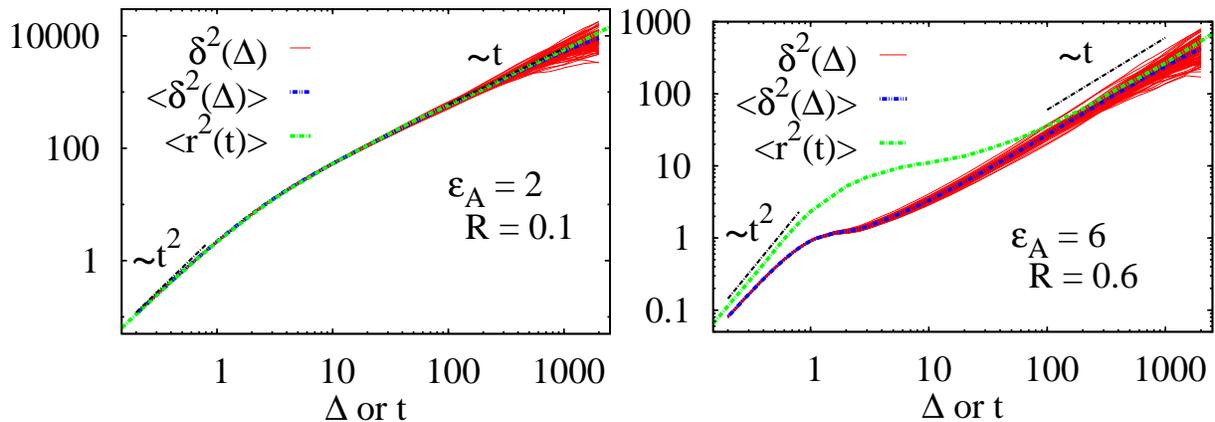}
\caption{MSD $\langle\mathbf{r}^2(t)\rangle$ (green curves) and time averaged MSD
$\overline{\delta^2(\Delta)}$ defined in Eq.~(\ref{eq-tamsd}) (red curves), as
well as the ensemble average $\left<\overline{\delta^2}\right>$ (blue curves)
over the simulated trajectories. Parameters are as indicated in the panels,
the number of traces used for averaging is $N=100$, and the length of the
trajectories is $T=10^4$ corresponding to $10^7$ simulation steps.}
\label{fig-tamsd}
\end{figure*}

\begin{figure}
\includegraphics[width=8cm]{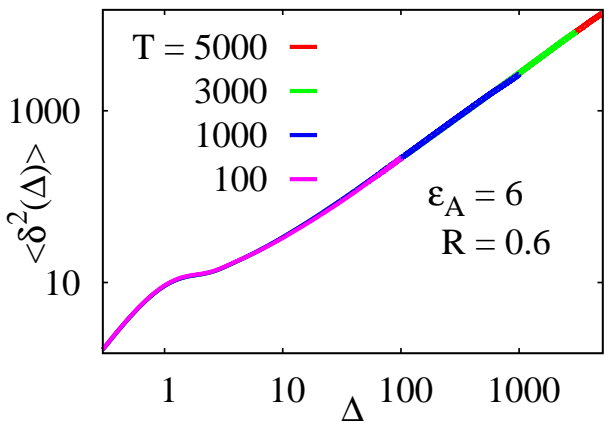}
\caption{Independence of the time averaged MSD $\left<\overline{\delta^2(\Delta)}
\right>$ on the length $T$ of the trajectory $\mathbf{r}(t)$. Parameters are the same as in Fig.~\ref{fig-tamsd}.} 
\label{fig-tamsd-t-varies}
\end{figure}

To obtain more insight into the characteristics of the diffusive motion of the
tracer particle in the crowded environment, we compute the time averaged MSD
\cite{franosch13,pccp,pccp1,pt,sokolov12}
\begin{equation}
\overline{\delta^2(\Delta)}=\frac{1}{T-\Delta}\int_0^{T-\Delta}\Big(\mathbf{r}
(t+\Delta)-\mathbf{r}(t)\Big)^2dt
\label{eq-tamsd}
\end{equation}
from the trajectory $\mathbf{r}(t)$ of the tracer of range $t=0,\ldots,T$. In
Eq.~(\ref{eq-tamsd}) $\Delta$ is the so-called lag time, which defines the size
of a window slid along the trajectory $\mathbf{r}(t)$. The length of the trajectory
$T$ is also referred to as measurement time. In addition to the individual time
traces $\overline{\delta^2(\Delta)}$ we also consider the average
\begin{equation}
\label{eatamsd}
\left<\overline{\delta^2(\Delta)}\right>=\frac{1}{N}\sum_{i=1}^{N}\overline{
\delta_i^2(\Delta)}
\end{equation}
over $N$ individual trajectories.
When measured time series are not long enough, this quantity provides smoother
curves if sufficiently many trajectories $N$ are available.

In Fig.~\ref{fig-tamsd} we show the time averaged MSD (\ref{eq-tamsd})
along with the MSD $\langle\mathbf{r}^2(t)\rangle$ of $N=100$ individual
trajectories. Let us first focus on the case of a moderate tracer-obstacle
attraction strength, $\epsilon_A=2k_B\mathcal{T}$. We observe that the individual
time traces $\overline{\delta^2(\Delta)}$ initially grow ballistically and then
cross over to normal diffusion, as indicated by the slopes. The results for
individual time traces $\overline{\delta^2(\Delta)}$ show only minute amplitude
variations at
different lag times, quantitatively similar to the spread of time traces of regular
Brownian particles. Of course, when $\Delta$ approaches the measurement time $T$,
the statistics of the time average defining $\overline{\delta^2(\Delta)}$ worsen
and some amplitude scatter occurs.
The average (\ref{eatamsd}) almost perfectly coincides
with the ensemble average MSD $\langle\mathbf{r}^2(t)\rangle$, compare the blue
and green curves in Fig.~\ref{fig-tamsd}. The latter observation corroborates the
ergodic nature of the tracer motion, that is, the equivalence of ensemble and long
time average of physical observables, here $\langle\mathbf{r}^2(\Delta)\rangle=
\left<\overline{\delta^2(\Delta)}\right>$ \cite{pt,pccp,pccp1}. The linear long
time scaling of the MSD defines the effective diffusion coefficient $D(\phi)$ we shown in
Fig.~\ref{fig-dc-ratio}.
Note that prolonged adsorption periods of the tracer on obstacle spheres correspond
to effective trapping and delays the growth of either $\left<x^2(\Delta)\right>$ or
$\overline{\delta^2(\Delta)}$. For more information on the
violation of the equivalence between time and ensemble averaged physical
observables in anomalous-diffusive stochastic processes we refer to the recent
review in Ref.~\cite{pccp1}.

As shown for the binding time statistics above, the
trapping times are exponentially distributed and thus the long time motion converges
to regular Brownian motion with a reduced, effective diffusivity $D(\phi)$. This
scenario is therefore fundamentally different from subdiffusive CTRWs \cite{montroll,report}, in which the characteristic trapping time
diverges. Our retarded Brownian motion-based physical rationale is consistent with 
experimental observations of protein diffusion in dense dextran solutions and with
Monte-Carlo simulations of tracer diffusion on lattices of immobile inert obstacles
as reported in Ref.~\cite{weiss09-fbm}. In addition, we checked that the average
time averaged MSD features no dependence on the trace length $T$: the values of
$\left<\overline{\delta^2(\Delta)}\right>$ almost perfectly overlap for different
trace-lengths $T$, as demonstrated in Fig.~\ref{fig-tamsd-t-varies}.

A different situation is encountered when we consider strong tracer-obstacle
attraction, $\epsilon_A=6k_B\mathcal{T}$ in Fig.~\ref{fig-tamsd}. We immediately
observe that up to $t=\Delta\approx10^2$ the time averaged MSD $\overline{\delta^2
(\Delta)}$ significantly differs from the corresponding ensemble average $\langle
\mathbf{r}^2(t)\rangle$. That is, on these time scales the systems exhibits the
disparity $\langle\mathbf{r}^2(\Delta)\rangle\neq\overline{\delta^2(\Delta)}$.
For times exceeding $t=\Delta\approx10^2$ the agreement between both quantities
becomes excellent, the system is asymptotically ergodic. Individual curves
$\overline{\delta^2(\Delta)}$ for single time traces show a somewhat increased
spread around their mean $\left<\overline{\delta^2(\Delta)}\right>$, however, this
is still within the range expected for (asymptotically) ergodic processes
\cite{jae_jpa}, and is significantly different from weakly non-ergodic processes
such as subdiffusive CTRWs \cite{he,pt,pccp,pccp1,sokolov12}
or heterogeneous diffusion processes \cite{hdp,pccp1}. For
diffusion processes of the subdiffusive CTRW type the spread of individual time
averaged MSD traces is expected to be \textit{finite\/} even for vanishing lag
times \cite{pccp,pccp1}. This kind of behaviour is definitely \textit{not\/}
observed in our simulations. Here a very narrow spread of $\overline{\delta^2}$
in the whole range of tracer-obstacle affinities and obstacle sizes is
observed, see, e.g., Fig.~\ref{fig-tamsd}. The transient non-ergodic
features observed here imply that the relaxation time towards ergodic behaviour is
increased for longer trapping times when the tracer-obstacle attraction is more
pronounced. The fact that the disparity $\langle\mathbf{r}^2(\Delta)\rangle\neq
\overline{\delta^2(\Delta)}$ is most pronounced around the turnover from initial
ballistic to terminal Brownian motion is consistent with observations of confined
stochastic processes driven by correlated Gaussian noise \cite{jae_pre,pccp1}.

We now address a quantity that is based on the fourth order moment of the time
trace $\overline{\delta^2(\Delta)}$. This non-Gaussianity parameter $G(\Delta)$
was shown to be a sensitive experimental indicator of the type of effective
stochastic process driving the tracer particle in crowded complex fluids
\cite{weiss14-pccp}. The non-Gaussianity parameter in three dimensions is
defined by \cite{franosch13} 
\begin{equation}
G(\Delta)=\frac{3}{5}\frac{\left<\overline{\delta^4(\Delta)}\right>}
{\left<\overline{\delta^2(\Delta)}\right>^2}-1.
\label{eq-g}
\end{equation}
For diffusion processes with a stationary Gaussian distribution of increments such
as Brownian motion and fractional Brownian motion we have $G=0$, while the parameter
$G$ becomes non-zero for processes with non-stationary increments and/or
non-Gaussian distributions such as subdiffusive CTRWs or
heterogeneous diffusion processes \cite{weiss14-pccp,hdp_pre}.

\begin{figure}
\includegraphics[width=8cm]{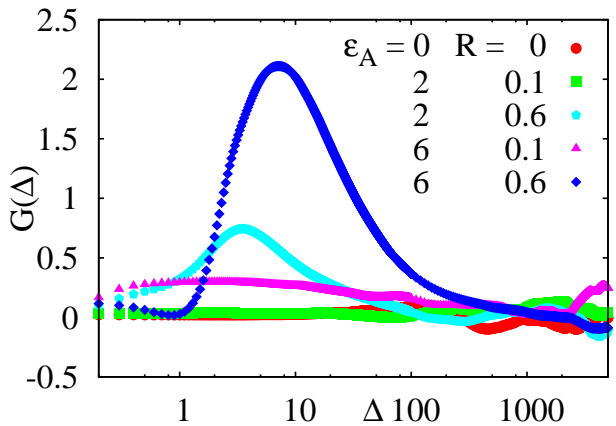}
\caption{Non-Gaussianity parameter (\ref{eq-g}) computed for the tracer diffusion
on a lattice of sticky obstacles, the parameters are indicated. As a reference for
almost perfect Gaussian behaviour $G=0$, we present the non-Gaussianity parameter
for vanishingly small, inert obstacles.
The length of individual simulated trajectories is
$T=10^4$.} 
\label{fig-non-g}
\end{figure}

We observe that for the simulated diffusion process in the presence of attractive
obstacles the non-Gaussianity parameter shown in Fig.~\ref{fig-non-g} is close to
zero for almost the entire length of the traces, apart from the initial regime of
the motion including inertial effects. These short time deviations from $G\approx0$
are particularly pronounced for relatively large obstacles with strong
tracer-obstacle attraction, as shown for the different parameters analysed in
Fig.~\ref{fig-non-g}. The fact that $G\approx0$ together with the equivalence of
the ensemble and time averaged MSDs at sufficiently long times are, of course,
a mirror for the Brownian nature of the observed motion. The smaller non-zero
values of $G$ detected in the long-time limit in Fig.~\ref{fig-non-g} are 
due to small discrepancies between the ensemble and time averaged MSDs.

At short to moderate times the non-Gaussianity parameter 
substantially deviates from the zero value characteristic of Brownian motion
for strongly adhesive and relatively large crowders, see, e.g., the blue curve
in Fig.~\ref{fig-non-g}. The deviations of $G(\Delta)$ occur at time scales of
$t\sim1\ldots30$ when the tracer diffusion is inherently non-ergodic, as we see
from the right panel of Fig.~\ref{fig-tamsd} plotted for the same parameters
($\epsilon_A=6 k_B\mathcal{T}$ and $R/R_{\text{max}}=0.6$). At very short times,
$t\ll1$, on which the MSD and time averaged MSD coincide, the non-Gaussianity
parameter respectively assumes very small values. Otherwise, the deviations
from $G=0$ we observe are within the range typically measured in experiments
\cite{weiss14-pccp} for asymptotically ergodic stationary-increment processes.

\section{Conclusions}
\label{sec-discussion}

We studied the passive motion of a tracer particle in an ordered, stationary
array of attractive spherical crowders. Based on a truncated attractive Lennard
Jones interaction between tracer and crowding obstacles, we simulate long
individual trajectories of the tracer based on the Langevin equation. The
resulting motion features a transition from initial ballistic flights
corresponding to the inertial particle motion to a long time Brownian diffusion
behaviour. The magnitude of the effective diffusion coefficient of this terminal
Brownian motion is reduced for increasing obstacle density and tracer-obstacle
attraction. However, a distinct non-monotonicity in this behaviour for large
obstacle radii and high attraction strength is observed, likely due to competing
attraction from multiple obstacles for which the tracer particle is in a limbo in
the bulk. The same effect also leads to a decrease of the time $t_A(\phi)$ spent
adsorbed to the obstacle surfaces during a long trajectory. The long time Brownian
dynamics was consistently shown to be associated with approximately vanishing
non-Gaussianity parameter. These dynamic features point out the crucial role of
varying tracer-obstacle binding strengths in the analysis of crowded systems, as
performed here. To put these qualitative statements on a more
physical foundation, additional simulations will be necessary. In particular,
we will study the time averaged van Hove cross-correlation functions
\cite{gel-no-go} for the tracer motion.

At intermediate times the tracer particle motion is anomalous, with a distinctly
time-dependent scaling exponent $\beta(t)$. In this regime the trapping to the
obstacles becomes the dominant mechanism. According to our results the time and
ensemble averaged MSDs are equivalent for moderate tracer-obstacle attraction
strength, however, a transient non-ergodic disparity between the two observables
is observed over a range of some $2.5$ orders of magnitude for stronger
tracer-crowder binding. This transient form
of weakly non-ergodic behaviour was previously observed for correlated Gaussian
processes under confinement \cite{jae_pre,lene1} of the otherwise ergodic process \cite{deng}. This behaviour should be kept in mind when precise diffusive
properties are to be analysed from measured or simulated time traces of our
system.

Extensions of the current model should consider a range of surface diffusivities
of tracer particles bound to obstacles. Moreover, the arrangement of crowders
should release the static, ordered arrangement on a lattice. Thus, off-lattice
simulations could include the motion of crowders around their equilibrium positions,
similarly to the analysis in Ref.~\cite{berry13}. Differences in the sizes of
individual crowders and a certain randomness in the tracer-obstacle affinity 
would be additional relevant generalisations of the current system. Mobile
obstacles were shown to profoundly shrink the time range of the transient
subdiffusive motion compared to immobile crowders \cite{berry13}. However, the
generality of these findings needs to be explored in a broader parameter range.
Currently, it remains elusive to arrive at realistic models capturing the
richness of real MMC in living biological cells with their wide variety of
crowder shapes, surface properties, persistence length and degree of branching
as well as a poly-disperse size distribution, in
addition to cellular structural elements, as well as charge effects
\cite{elcock-ecoli-cytoplasm}. Finally, active processes such as energy-consuming
transport in living cells \cite{elbaum,robert,fran13-driven} needs to be added to
achieve a closed picture of all facets of cellular dynamics.

\renewcommand{\thefigure}{A\arabic{figure}}

Our study complements several other recent analyses of tracer motion in crowded
environments. Thus, tracer diffusion in a system of relaxed and stretched polymer
chains in the presence of tracer-polymer attraction was studied by Langevin
dynamics simulations with the Espresso package \cite{holm11-polymers}. We observe
that such obstructed diffusion with sticky obstacles resembles our current
results. For instance, the evolution of the time-dependent MSD scaling exponent
$\beta$ (see Fig.~4 in Ref.~\cite{holm11-polymers}) shows a transition from the
initial ballistic regime to a subdiffusive regime at intermediate times, and
further to Brownian motion in the long-time limit. The anomalous diffusion regime
was shown to span a larger time window for the relaxed as compared to the stretched
self-avoiding chains \cite{holm11-polymers}. Higher polymer densities and stronger
tracer-polymer interactions yield wider regions of tracer subdiffusion
\cite{holm11-polymers}. A continuation of this study for tracer diffusion in a
cylindrical pore with surface-grafted polymer brushes of varying density showed
that the subdiffusive regime is more pronounced for weak-to-moderate tracer-polymer
interaction \cite{holm13-cylinder}.

We also mention an experimental study of
impeded colloidal diffusion in transient polymer networks with varying
colloid-polymer binding interactions \cite{colloid-diffusion-polymers-exper}.
For the diffusion of an inert tracer in a responsive elastic network system,
when the tracer size is of the same order as the unit cell of this gel, transient
subdiffusion was reported and shown to involve characteristic collective dynamics
of tracer and gel \cite{gel-no-go}. The transient subdiffusion in this study is
in contrast to the experimentally observed long-tailed distribution of trapping
times of sub-micron tracers in semi-flexible, inherently dynamic networks of
cross-linked actin
\cite{bausch-actin-subdiffusion} and thus further underlines the non-universal
character of the dynamics in crowded systems. We finally mention that diffusion
in two-dimensional, oriented fibrous networks in the presence of repulsive and
attractive particle-obstacle interactions was in fact studied experimentally in
connection with hydrogel-like structures of the extracellular matrix
\cite{bausch-es-fibers}.

Is crowding in cells merely an effect of cramming a rich multitude of different
bio-molecules into a minimal volume, or does it have an evolutionary purpose in
giving rise to dynamic phenomenon such as (transient) subdiffusion \cite{goldingcox,
guigas}? The combination of advanced single particle technology
and other experimental methods along with improved in silico studies will lead
to significant advances in the understanding of these still elusive questions.

\section{Acknowledgements}

The authors acknowledge funding from the Academy of Finland (FiDiPro scheme to
RM), the German Research Foundation (DFG Grant CH 707/5-1 to AGC), and the
German Ministry of Education and Research (BMBF Grant to SG).





\begin{thebibliography}{99}
\bibitem{note-berry} Moreover, in Ref.~\cite{berry11} the
distribution of particle trapping times in dynamical cages formed between the
crowders was shown to be inconsistent with the CTRW model, as observed in
simulations for mobile crowders, crowders confined in a potential, and static
obstacles.  Namely, the power-law distribution of trapping times $\psi(\tau)\sim
\tau^{-\gamma}$ was shown to range between $\gamma=3.5\dots4$, in contrast to
the range $\gamma=1\dots2$ expected for subdiffusive CTRW processes with
diverging characteristic waiting time \cite{pt,pccp,pccp1}.
\end{thebibliography}
\end{document}